# Can We Aggregate Human Intelligence? An Approach for Human Centric Aggregation using Ordered Weighted Averaging Operators

Shahab Saquib Sohail[1,*], Jamshed Siddiqui[2], Rashid Ali[3], S. Hamid Hasan[4] and M.Afshar Alam [5]

[1,5] Department of Computer Science and Engineering, Jamia Hamdard, New Delhi, India
[1,*]shahabssohail@jamiahamdard.ac.in, [4]aalam@jamiahamdard.ac.in,

[2]Department of Computer Science, Aligarh Muslim University, Aligarh, India
jamshed_faiza@rediffmail.com

[3]Department of Computer Engineering, Aligarh Muslim University, Aligarh, India
rashidaliamu@rediffmail.com

[4]Faculty of Computing and Information Technology, King Abdul Aziz University, Jeddah, KSA.
shhasan@kau.edu.sa

**Abstract:**

The primary objective of this paper is to present an approach for recommender systems that can assimilate ranking to the voters or rankers so that recommendation can be made by giving priority to experts' suggestion over usual recommendation. To accomplish this, we have incorporated the concept of human-centric aggregation via Ordered Weighted Aggregation (OWA). Here, we are advocating ranked recommendation where rankers are assigned weights according to their place in the ranking. Further, the recommendation process which is presented here for the recommendation of books to university students exploits linguistic data summaries and Ordered Weighted Aggregation (OWA) technique. In the suggested approach, the weights are assigned in a way that it associates higher weights to best ranked university. The approach has been evaluated over eight different parameters. The superiority of the proposed approach is evident from the evaluation results. We claim that proposed scheme saves storage spaces required in traditional recommender systems as well as it does not need users' prior preferences and hence produce a solution for cold start problem. This envisaged that the proposed scheme can be very useful in decision making problems, especially for recommender systems; in addition, it emphasizes on how human-centric aggregation can be useful in recommendation researches, and also it gives a new direction about how various human specific tasks can be numerically aggregated.

**Introduction:**

The recent research in decision making has explored new directions of natural language usuality which has given an emergence to a new perspective of information aggregation that can be termed as 'human-centric aggregation'. The primary objective of this paper is twofold. First, to incorporate the concept of human-centric aggregation via Ordered Weighted Aggregation (OWA) which is recently presented by J. Kacprzyk, R. Yager and J.M. Merigo in the memorial issue of the reputed IEEE computational Intelligence magazine[1]. Second, to design a recommender systems that can assimilate ranking to the voters or rankers and assign them weights accordingly which in turn may produce recommendation where experts suggestion is given priority over usual recommendation. It is interesting to explore how the human-centric aggregation can be useful in solving real life problem. By human-centric aggregation we mean quantifying those qualitative attributes which are human specific like, judgement, intelligence, intention, and vision etc. The OWA has been extensively used in literature for exploring different aspects of human specific problems [2-6], especially in decision making problems. We have tried to explore its diversity and strength for recommender systems. Although, OWA has been used in recommendation previously; we have modified the weight assignment method for it. The new weight assignment method gives more weightage to the best ranked. We use 'most preferred first' linguistic quantifier with OWA, for the purpose.

In this paper, we are intended to recommend top books to universities 'students by aggregating the suggestion of experts from top ranked institutions. The idea is supported from previous work which uses OWA for book recommendation, however, priority to best ranked university or to experts of the subjects is not adequately provided there [7]. The issue lies with weight assignment with some of the fuzzy linguistic quantifiers which generates zero values to institution which is top ranked. Therefore, we have assigned weights in a way that best ranked university is achieved higher weights in comparison to lower ranked universities. The proposed scheme will also save the space as it does not need to record their prior preferences, which most of the existing recommender systems do.

The results of proposed OWA (most preferred first) is compared with previous positional aggregation based scoring (PAS) technique and OWA with other linguistic quantifiers on 8 different parameters. The results show that proposed scheme has about 17% improved performance with compare to PAS and up to 65% with compare to ordered weighted aggregation where weights are not assigned according to the order of ranking. Further, it is elaborately discussed how the OWA can be perceived as human-centric aggregation. This envisaged that the proposed scheme can be very useful in various decision making problems, especially for recommender systems; in addition, it gives new direction about how various human specific task can be numerically aggregated.

In section 2, perspective of human-centric aggregation including back ground of OWA is discussed. Section 3 explains the proposed scheme of recommendation of books using OWA operators. In Section 4, detail discussion of experiments, dataset and results are given, further, performance evaluation strategy is illustrated with diagram. Finally we conclude in section 5.

### 1. Human-Centric Aggregation:

In our daily life we have come across number of issues where human perception plays an important role without which decision making becomes difficult. It is obvious that human being is the core for aggregating different opinion to come on one conclusion [8,9]. The example includes all those problems where consensus is required, like decision making, voting results, etc. in the same way, the aggregation of numerical values is important. R. Yager has significantly contributed to the science of aggregation by introducing Ordered Weighted Aggregation (OWA) [4]. We describe OWA in the following section.

#### 1.1. Ordered Weighted Aggregation (OWA)

OWA has been used extensively in the literature especially to handle the uncertainty [8-15]. The authors have used various OWA based applications which include techniques for randomized queries for searching over web [16,17], applying aggregation operators based on fuzzy concepts for recommender systems [18], social networking[19], GIS applications [12] and environments [20-22], and combination of OWA & opinion mining for book recommendation [23]. It is also used for analysing talent and skills of the players for different sports [2]. The frequent use of OWA in multi criteria decision making has been reported in [24-30] whereas the fuzzy methods are supposed to be very impressive [31-33] in decision making problems too.

Ordered weighted aggregation (OWA) can be termed as function from $R^n \rightarrow R$, where 'W'- weight vectors is associated to them in such a way that $\sum_{k=1}^{n}(W_k) = 1$ and $W_k \in [0,1]$. Mathematically it is given as;

$$OWA\ (d_1, d_2, \ldots, d_n) = \sum_{k=1}^{n}(W_k C_k) \quad \text{-----------(1)}$$

where, if we sort $C_k$, $k^{th}$ largest element would be $d_k$. we are primarily intended to incorporate human-specific aggregation for recommendation process (books) with the help of OWA, therefore, the point that must be addressed is about how many books to recommend, how many users are needed to be involved, some, almost all, most, etc. Therefore, we use linguistic quantifiers. For fuzzy linguistic quantifier, we define Function Q(r) for relative quantifier as:

$$Q(r) = \begin{cases} 0 & \text{if } r<a \\ \frac{(r-a)}{(b-a)} & \text{if } a \leq r \leq b \\ 1 & \text{if } r>b \end{cases} \quad \text{--- (2)}$$

where Q (0) = 0, $\exists r \in [0, 1]$ such that Q(r) =1, and a, b $\in$ [0,1].

for above condition, we have Q: [0, 1] → [0, 1].

The weights '$W_k$' for OWA operator is calculated by following equation [4,17].

$$W_k = \{Q\ (k/m) - Q\ ((k-1)/m)\}, \quad \text{-----------(3)}$$

where k = 1, 2… m.

Different weights can be obtained with the help of different linguistic quantifier. Like, for 'Most' linguistic quantifier, a=0.3 and b=0.8, by the use of these quantifiers those books are preferred which are recommended by most of the universities.

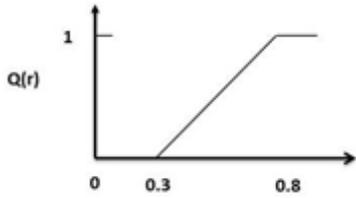 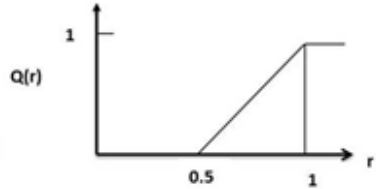 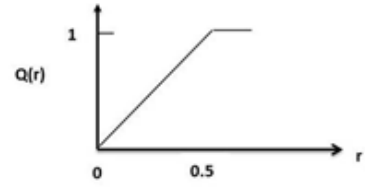

Figure 1(a) most quantifier

Figure 1(b) as many as possible quantifier

Figure 1(c) at least half quantifier

Similarly, 'as many as possible' and 'at least half' are other quantifiers for which values of (a, b) are (0.5, 1) and (0, 0.5) respectively. Graphical representations of these fuzzy linguistic quantifiers are shown for 'most', 'as many as possible', and 'at least half', respectively. Figure 1(a), (b) and (c) represent graph for the above quantifiers.

Now, considering a situation when we need to rank the voters, i.e. rankers are valued and they influence decision making. We suggest OWA with a modification in the weight assignment of equation (3), and term it as OWA (most preferred first). For this, we define weight assignment as –

$$W_k = \frac{u+1-k}{N} \quad \text{----------------------------- (4)}$$

Where, total no. of universities is given by 'u'; $N = \sum_{i=1}^{u}(k)$. Also, $W_k \in [0,1]$ and $\sum_{1}^{u} W_k = 1$.

OWA (most preferred first) is given as -

$$OWA\ (most\ preferred\ first) = \sum_{k=1}^{n}(W_k C_k) \quad \text{-------------------------------- (5)}$$

The application of the above suggested concept can influence human-specific problems where rankers or experts are needed to be assigned weights, i.e., recommendations are given by the experts. These problems may have wide domain including judgement, human intelligence, voting results, and any problem that involves consensus.

## 2. Proposed Recommendation Strategy using OWA (most preferred first)

Since, we are intended to gives higher weights to best ranked entity. Therefore, if $W_k > W_m$ when k<m. i.e. for three ordered ranked universities Univ_1, Univ_2, and Univ_3, we must have $W_1 > W_2 > W_3$. Here, we are aimed at proposing recommendation strategy for books. Thus, we claim, it is important to know the experts ranking and the suggested technique represented by equation 5 can even weights those experts too, hence generating more appropriate recommendation. Keeping this concept in view, we proceed for recommendation.

Since, we are intended to recommend books for computer science under-grad students. Initially, all the possible books have been incorporated for the purposes of experiment. But, without any limit and criteria, it would have a huge data and wastages of storage. Therefore, we filtered the data by applying PAS concept [34] and incorporated books from top universities as prescribed in their syllabus. The top Indian University from QS ranking [35] has been taken and top Indian institutes which comes in world top ranking, are being examined for our work. Only 'computer science' are taken as a subject of interest as we just intended to show how the human-centric aggregation can perform, once it can happen with a smaller dataset, it can also be easily extended to larger data set. Hence, courses of computer science like, computer network, database concepts, etc. are searched in these institutions and their recommended books are stored. Only 7 Indian Universities got position in QS world ranking, which are listed in Table 1.

**Table 1: Top ranked seven Universities of India in QS ranking [35].**

| Rank Position | University Name |
|---|---|
| 1 | IIT, Bombay |
| 2 | IIT, Delhi |
| 3 | IIT, Kanpur |
| 4 | IIT, Madras |
| 5 | IISC, Bangalore |
| 6 | IIT, Kharagpur |
| 7 | IIT, Roorkee |

The positional aggregation based scoring (PAS) technique has been used to quantify the ranking. Which assign a numerical value corresponding to the rank of the books, for details please see [36].

The PAS technique tries to assign a maximum value to best ranked university, and quantifies every ranking to a numerical value. OWA based approach for Book Recommendation (most preferred first) is shown in Figure 2. Once we find the positional score for books, we assign weights for ranked

Universities. We use equation (4) and (5) for finding final score of books. The sorted value gives the books ranking.

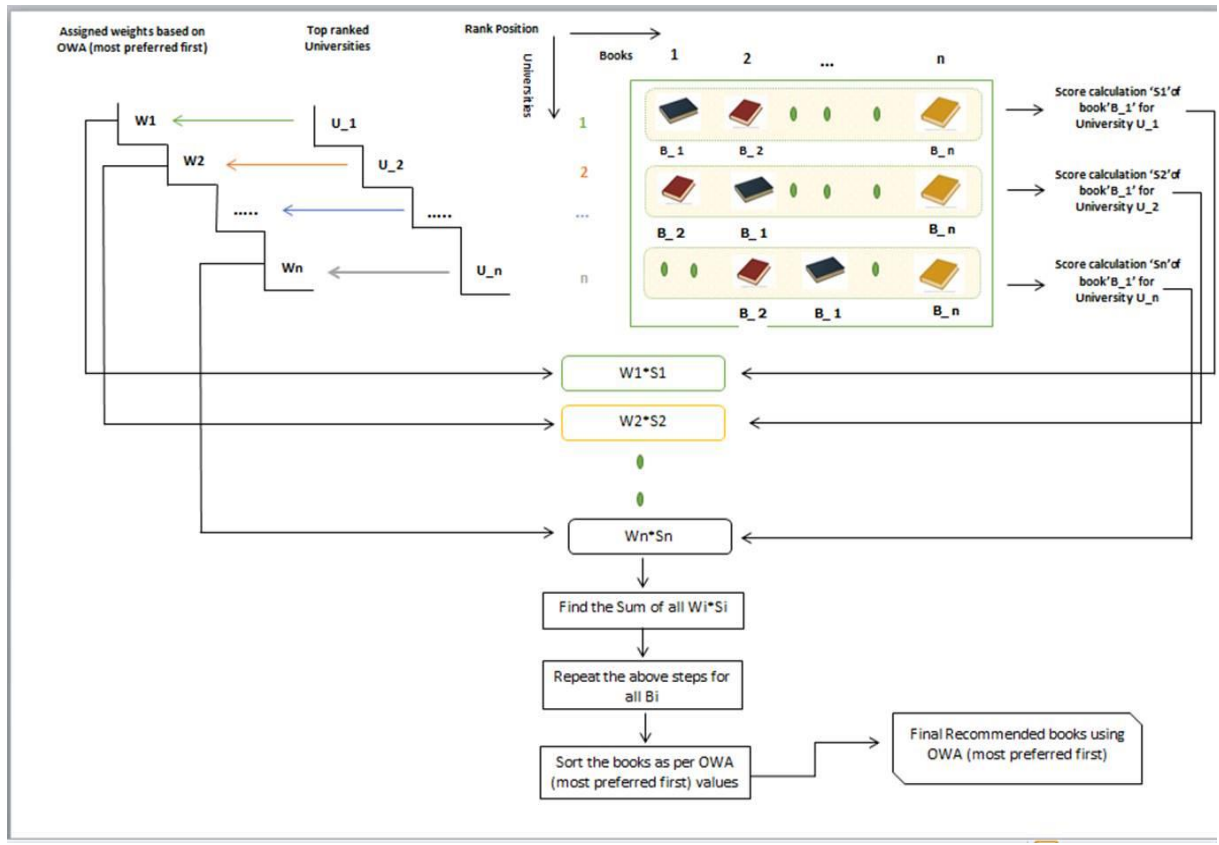

**Figure 2. OWA based approach for Book Recommendation (most preferred first)**

**Table 2: List of computer science' courses which have been included in the Syllabus at top Universities.**

| Sequence | Course Title | Univ_1 | Univ_1 | Univ_1 | Univ_1 | Univ_1 | Univ_1 | Univ_1 |
|---|---|---|---|---|---|---|---|---|
| | Artificial Intelligence | 👍 | 👍 | 👎 | 👎 | 👍 | 👍 | 👍 |
| | Compiler Design | 👍 | 👍 | 👍 | 👎 | 👍 | 👎 | 👍 |
| | Computer Networks | 👍 | 👍 | 👍 | 👎 | 👍 | 👍 | 👍 |
| | Discrete Mathematics | 👍 | 👍 | 👎 | 👍 | 👍 | 👍 | 👍 |
| | Data Structure | 👍 | 👍 | 👍 | 👍 | 👍 | 👍 | 👍 |
| | Graphics | 👎 | 👎 | 👎 | 👍 | 👍 | 👍 | 👍 |
| | Operating Systems | 👍 | 👍 | 👎 | 👎 | 👍 | 👍 | 👍 |
| | Principles of Database Systems | 👍 | 👍 | 👍 | 👍 | 👍 | 👍 | 👍 |
| | Software Engineering | 👍 | 👍 | 👍 | 👎 | 👍 | 👍 | 👍 |
| | Theory of Computation | 👍 | 👍 | 👎 | 👎 | 👍 | 👍 | 👍 |

Only those courses which are kept in the curriculum by these leading institutions have been taken into consideration for the experiment. A complete list is given in Table 2. Where '✎' shows the course is available and '✏' indicates non-availability.

The universities usually display the recommended books on their websites, and we have sincerely tried to fetch all those books and categorize distinct book differently. For some courses for which books by some particular universities are not displayed, we have contacted in person and by email to gather the required information. Ten (10) different courses have been added in Table 2. It is evident from the table, not all university has published books on a topic at their respective web sites, at the same time not every book has been recommended by all the university. As a result, we have gathered 158 different books for above 10 courses which have been included in experimental procedure. The process of selecting books only from ranked universities serves the purpose of reducing huge amount of the related books available, which makes the procedure easier.

## 3. Results and Discussions

*3.1 Evaluation Metrics:*

There are different metrics for evaluations of the recommender systems which have been used in the literature [37-42]. Some of them are veracity measure, i.e. they are used to measure accuracy. The more the value of the measure, the better is the result. On the other hand some of them are fallacy measure, i.e. the minimum resulting value indicates the better performance. We have used four veracity measures and four fallacy measures in our results for deciding the performance of the proposed mechanism. For measures of precision we have used p@10, Mean Average Precision (MAP), Mean Reciprocal Rank (MRR), and Modified Spearman Rank Correlation Coefficient (MSRCC). whereas for the parameters to be termed as fallacy measures, we have used FPR@10, FNR@10, Root Mean Square Error (RMSE) and Mean Absolute Error (MAE). The list and details of the evaluation metrics are given below.

  i. P@10
 ii. FPR@10
iii. FNR@10

iv.   Mean Average Precision (MAP)

   v.   Mean Absolute Error (MAE)

   vi.   Mean Reciprocal Rank (MRR)

   vii.   Root Mean Square Error (RMSE)

   viii.   Modified Spearman rank Correlation Coefficient (MSRCC)

### 3.1.1 P@10

We denote the precision at top-10 positions as P@10 and define it for our purpose as;

$$P@10 = \frac{Number\ of\ books\ endorsed\ by\ user\ as\ well\ as\ recommended\ in\ top\ 10\ position}{10} \quad \text{--- (6)}$$

P@10 is obtained by comparing the ranking that comes out by applying OWA (most preferred first) techniques with the experts ranking. These values are shown in Figure 3(a).

### 3.1.2 FPR@10

FPR@10 denotes false positive rate for top 10 positions which is defined as follows:

$$FPR@10 = \frac{Number\ of\ books\ that\ comes\ in\ top\ 10\ position\ which\ is\ not\ liked\ by\ customer}{10} \quad \text{--- (7)}$$

The "false positive" is a case when recommended items are different from users' preferred items. This situation leads to irritation of the customer and hence treated as the worst scenario. It may cause to damage in business as it may block the customers from further buying.

FPR@10 is obtained by the comparison of the ranked-position that comes out by applying OWA (most preferred first) techniques with the experts ranking. These values are shown in Figure 3(b).

### 3.1.3 FNR@10

The false negative rate is the error that gives the idea about a situation when the recommendation technique misses out to recommend items which are preferred by customers. The false negative rate for top 10 positions is denoted as FNR@10. In the context of our problem, we define it as follows:

$$FNR@10 = \frac{Number\ of\ books\ which\ is\ not\ recommended\ but\ liked\ by\ customers\ in\ top\ 10}{10} \quad \text{--- (8)}$$

FNR@10 is shown in Figure? These values are obtained by comparing the ranking that comes out by applying OWA with linguistic quantifier techniques with the experts ranking.

### 3.1.4 Mean Average Precision

Mean Average Precision (MAP) is mathematically defined as;

$$\text{Mean Average Precision} = (1/n) \sum_{i=0}^{n} p(C_i) \quad \text{--- (9)}$$

$p(C_i)$ represent precision for $i^{th}$ customer where 'n' is the total number of customers concerned in the experiment.

### 3.1.5 Mean Absolute Error

The Mean Absolute Error (MAE) tells how close the outcome with actual result is. It is given by;

$$\text{Mean Absolute Error} = (1/n) \sum_{i=1}^{n} (|O_i - A_i|) \quad \text{---- (10)}$$

In the above equation – (10) the observed values and actual values are represented as $O_i$ and $A_i$ respectively where n symbolizes the number of observation.

### 3.1.6 Mean Reciprocal Rank

Let 'r' denotes the rank of a product in the proposed approach (which is based on OWA (most preferred first) technique). In the ranking of the products as suggested by the proposed scheme in the paper, we aimed at finding the ranked-position of an item when it is known that it has ranked first in the system ranking. Reciprocal Rank (RR) is given as:

$$RR = \frac{1}{r}$$

Mean Reciprocal Rank (MRR) is calculated for first ranked product of all the items. Mathematically it is given by;

$$\text{Mean Reciprocal Rank} = (1/n) \sum_{i=1}^{n} (RR_i) \quad \text{- - - - (11)}$$

Where 'n' represents total number of items and 'i' represents $i^{th}$ items, respectively. Mean Reciprocal Rank (MRR) measures how relevant is the product for a customer as it suggests the best item. If the position of first ranked item by the experts as well as by the proposed scheme coincides, it simply indicates the item is of great interest. Since, the MRR comes out to be 1, which implies best case.

*3.1.7 Root Mean Square Error*

The root mean square error is used to measure error value. It is defined as;

$$Root\ Meam\ Square\ Error = \sqrt{(1/n) \sum_{k=0}^{n}(Y_i - y_i)^2} \ \ ----(12)$$

$Y_i$ and $y_i$ indicate two different entities, one represents actual ranking whereas another stands for outcomes of the ranking by experiments. In the above equation, actual ranking is basically expert's recommendation which is denoted by '$Y_i$' whereas '$y_i$' denotes system's prediction.

*3.1.8 Modified Spearman rank Correlation Coefficient*

The modified spearman rank correlation coefficient is suggested after the spearman correlation coefficient proved incapable of producing correct result for partial list. The mathematical definition of the modified spearman rank correlation coefficient is given as-

$$rs' = 1 - \frac{\sum_{i=1}^{m}(i-V_i)^2}{m\,([\max\{V_j\}_{j=1}^{m}]^2-1)} \ \ ------(13)$$

where full list and partial list is given by [1, 2… m] and [$v_1$, $v_2$… $v_m$] respectively.

### 3.2 Experimental Results

The methods discussed in the section 3 are illustrated here. Since, the procedure is same for all the books of each course; therefore, we have demonstrated the examples considering only books of one course only. For the sake of simplicity, the books on 'Data structure' are considered. As different ranked university has different ranking of books in which there may be same book repeated everywhere, or may be only one book which is included in the ranking of the respective universities. The respective university wise ranking of the books on data structure is listed in Table 3. The books on Data Structure (DS) are represented by code, 'DS1', 'DS2' etc. we can easily find that the book DS1 is placed in 1st rank by Univ_1. At the same time no two universities has recommended same book on first rank, even no two books have been included in the ranking of more than two universities. Only books DS1 and DS3 are included twice in top ranked universities ranking.

**Table 3: Ranking of books on 'Data Structure' by top universities.**

| Rank position | U₁ | U₂ | U₃ | U₄ | U₅ | U₆ | U₇ |
|---|---|---|---|---|---|---|---|
| 1st | DS1 | DS2 | DS4 | DS9 | DS12 | DS9 | DS15 |
| 2nd | x | DS3 | DS5 | DS1 | DS8 | DS14 | DS16 |
| 3rd | x | x | DS6 | DS10 | DS13 | DS10 | x |
| 4th | x | x | DS7 | DS11 | DS3 | x | x |
| 5th | x | x | DS8 | x | x | x | x |
| 6th | x | x | x | x | x | x | x |
| 7th | x | x | x | x | x | x | x |
| 8th | x | x | x | x | x | x | x |
| 9th | x | x | x | x | x | x | x |
| 10th | x | x | x | x | x | x | x |

A total of 16 books on data structure have been included in the ranking of top 7 universities where the first ranked university has only one book in their prescribed syllabus. The Positional aggregation technique based score has been obtained by the procedure as stated[34] and discussed in section 3. For the sake of simplicity and to save the space, we have represented final score of the books of one course (data structure) only, which is given in Table 4. Here, in the paper we are intended to show how the human centric aggregation can perform, and how the change in weight assignment will enhance the aggregation for a problem which incorporate human-specific decisions. Therefore the results which have been obtained by applying OWA (most preferred first) are shown. The method for a single book is illustrated below. Here we have termed criteria as selection of books by a university; therefore we have classified seven criteria of selection, which is basically total number of universities under consideration. The weight assignment formula as suggested in equation (4) is used to calculate weights for seven (7) criteria. The weights for OWA (most preferred first) are given in Table 5.

Further, the existing strategies have been compared with the proposed mechanism which will be discussed in the subsequent section. The final values for positional score, OWA (at least half), OWA (as many as possible), OWA (most) and OWA (most preferred first) for books on data structure (DS) is shown in Table 6. The tabulated score tells which method to be reckoned as the best for the books concerned.

The procedure for calculation of OWA (most preferred first) using positional score is as follows:

Let the positional scores (for DS1) are represented as:

$$C_k = \begin{pmatrix} 1 \\ 0 \\ 0 \\ 0.9375 \\ 0 \\ 0 \\ 0 \end{pmatrix}$$

**Table 4: Quantified Positional Score for books on Data Structure.**

| Book Code | $U_1$ | $U_2$ | $U_3$ | $U_4$ | $U_5$ | $U_6$ | $U_7$ |
|---|---|---|---|---|---|---|---|
| **DS.1** | 1 | 0 | 0 | 0.9375 | 0 | 0 | 0 |
| **DS.2** | 0 | 1 | 0 | 0 | 0 | 0 | 0 |
| **DS.3** | 0 | 0.9375 | 0 | 0 | 0.8125 | 0 | 0 |
| **DS.4** | 0 | 0 | 1 | 0 | 0 | 0 | 0 |
| **DS.5** | 0 | 0 | 0.9375 | 0 | 0 | 0 | 0 |
| **DS.6** | 0 | 0 | 0.875 | 0 | 0 | 0 | 0 |
| **DS.7** | 0 | 0 | 0.8125 | 0 | 0 | 0 | 0 |
| **DS.8** | 0 | 0 | 0.75 | 0 | 0.9375 | 0 | 0 |
| **DS.9** | 0 | 0 | 0 | 1 | 0 | 1 | 0 |
| **DS.10** | 0 | 0 | 0 | 0.875 | 0 | 0.875 | 0 |
| **DS.11** | 0 | 0 | 0 | 0.8125 | 0 | 0 | 0 |
| **DS.12** | 0 | 0 | 0 | 0 | 1 | 0 | 0 |
| **DS.13** | 0 | 0 | 0 | 0 | 0.875 | 0 | 0 |
| **DS.14** | 0 | 0 | 0 | 0 | 0 | 0.9375 | 0 |
| **DS.15** | 0 | 0 | 0 | 0 | 0 | 0 | 1 |
| **DS.16** | 0 | 0 | 0 | 0 | 0 | 0 | 0.9375 |

**Table 5: Weights assigned to Universities by using OWA (most preferred first)**

| Ranked University | Weights assigned |
|---|---|
| Univ_1 | W1=0.25 |
| Univ_2 | W2=0.21428 |
| Univ_3 | W3=0.17857 |
| Univ_4 | W4=0.14285 |
| Univ_5 | W5=0.10714 |
| Univ_6 | W6=0.07142 |
| Univ_7 | W7=0.03571 |

These values are reordered and represented in descending order. Let the order is $d_k$. The re-ordered value shall be represented as-

$$d_k = \begin{pmatrix} 1 \\ 0.9375 \\ 0 \\ 0 \\ 0 \\ 0 \\ 0 \end{pmatrix}$$

From equation (4) and (5), we get –

$$OWA \ (most \ preferred \ first) = OWA(d_1, d_2, \ldots, d_n) = \sum_{k=1}^{n}(W_k C_k)$$

Therefore we obtain $OWA \ (d_1, d_2, \ldots, d_n)$ as:

$$= [0.25, \ 0.21428, 0.17857, 0.14285, 0.10714, 0.07142, 0.03571] \times \begin{pmatrix} 1 \\ 0.9375 \\ 0 \\ 0 \\ 0 \\ 0 \\ 0 \end{pmatrix}$$

$= (0.25 \times 1 + 0.21428 \times 0.9375 + 0.17857 \times 0 + 0.14285 \times 0 + 0.10714 \times 0 + 0.07142 \times 0 + 0.03571 \times 0)$

$= 0.450813$

**Table 6: Final scores and ranked list of books on data structure using OWA (most preferred first)**

| Rank position | Book Code | Score obtained using OWA (most preferred first) techniques |
|---|---|---|
| 1st | DS.9. | 0.4642 |
| 2nd | DS.1. | 0.450813 |
| 3rd | DS.3. | 0.408413 |
| 4th | DS.10. | 0.406175 |
| 5th | DS.8. | 0.395025 |
| 6th | DS.4. | 0.25 |
| 7th | DS.12. | 0.25 |
| 8th | DS.15. | 0.25 |
| 9th | DS.5. | 0.234375 |
| 10th | DS.14. | 0.234375 |
| 11th | DS.16. | 0.234375 |
| 12th | DS.6. | 0.21875 |
| 13th | DS.13. | 0.21875 |
| 14th | DS.2. | 0.2142 |
| 15th | DS.7. | 0.203125 |
| 16th | DS.11. | 0.203125 |

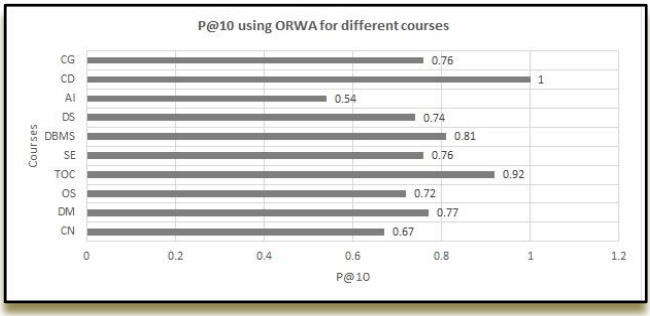

**Figure 3 (a) P@10 using OWA (mot preferred first) for books of all courses**

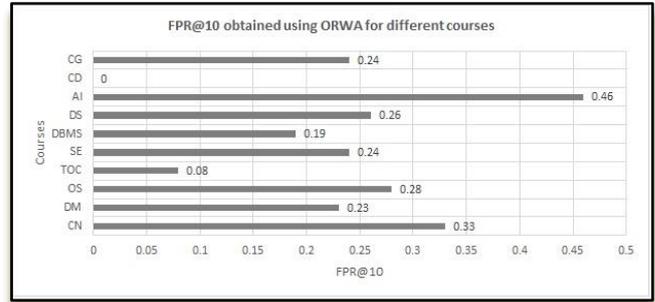

**Figure 3 (b) FPR@10 using OWA (mot preferred first) for books of all courses**

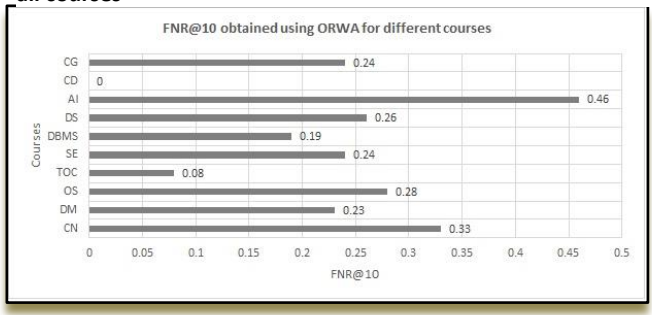

**Figure 3 (c) FNR@10 using OWA (mot preferred first) for books of all courses**

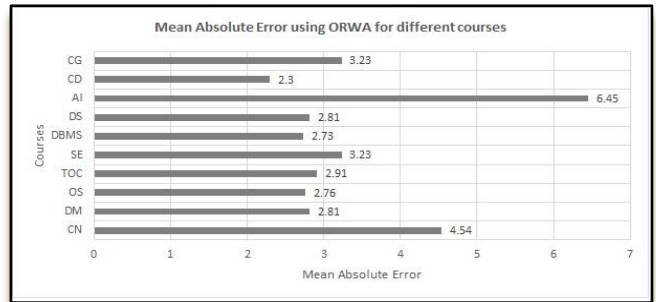

**Figure 3 (d) Mean Absolute Error using OWA (mot preferred first) for books of all courses**

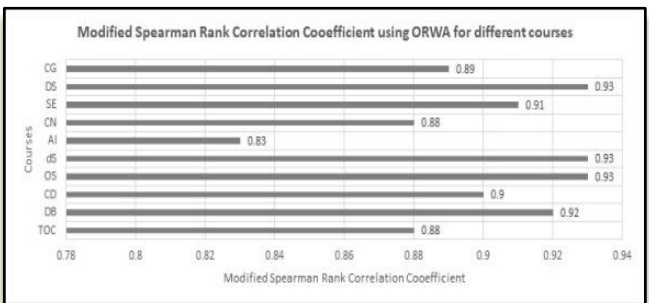

**Figure 3 (e) Modified Spearman Rank Correlation coefficient using OWA (mot preferred first) for books of all courses**

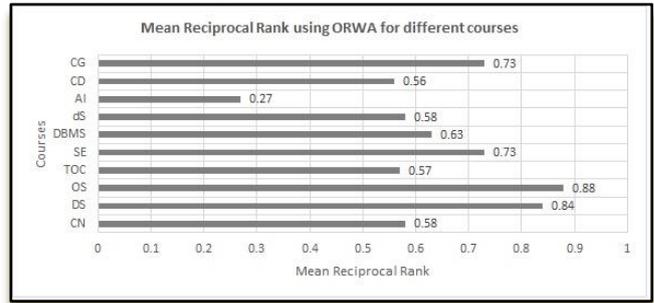

**Figure 3 (f) Mean Reciprocal Ranking using OWA (mot preferred first) for books of all courses**

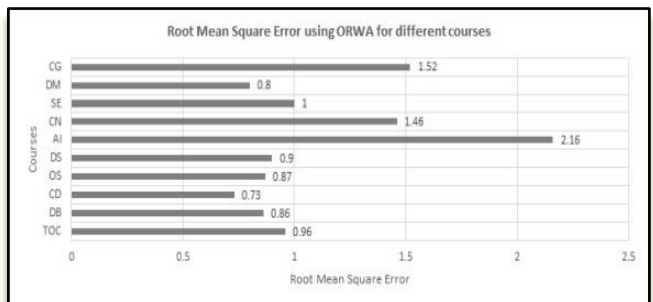

**Figure 3 (g) Root Mean Square Error using OWA (mot preferred first) for books of all courses**

**Figure 3. Results of different parameters using OWA (mot preferred first) for books of all courses**

These scores are calculated for all the books, these scores help in sorting the books which provides a platform for ranking of the books for above method. The final OWA scores with ranking of the respective books for the linguistic quantifier 'most preferred first' are shown in Table 6.

The OWA (mot preferred first) score comes out be 0. 450813 for DS1 as shown in section 4.2. Similarly values for all books have been calculated. With the help of the proposed mechanism the above scores are calculated which lays a foundation to the ranking of the books by different approaches, the ranking for books on DS for OWA (mot preferred first) is given in Table 6. In the same way we have calculated ranking by all related methods of all books for different courses. In the comparison section all these rankings are considered but we have not shown those scores and ranking here in the table to save the space and avoid the repetition. Thus there are total 158 different books have been filtered from huge amount of books in the process of recommendation, which eases the complexities of the recommendation process. All the related approaches namely PAS, OWA (at least half), OWA (as many as possible) and OWA (most) have been tested for the same number of books and all these values are stored. The OWA with quantifier 'mot preferred first' performance for all the books of all the courses with respect to eight different parameters is shown in Figure 3.

**3.3 Performance evaluation mechanism**

The above methods which have been discussed in the preceding sections, explain the way books are recommended for students on a particular subject. The QS ranking[35] is considered as the base of the complete recommendation process. The books recommended at the top ranked Indian Institute at QS world university ranking are taken into account for experiments. Several techniques have been applied previously to filter the huge record for recommending most promising books to readers. These techniques include PAS technique, OWA (at least half), OWA (as many as possible) and OWA (most). These techniques are compared here with the proposed methodology, that incorporates OWA with modified weight assignment formula, which we say is more human-centric and we term it as OWA (mot preferred first). Since PAS technique does not involve any weights to be assigned in its process of recommendation, thus, we say the process is an un-weighted aggregation. The scores are calculated using PAS technique, these scores are assigned fuzzy weights using OWA [(at least half), (most) and (as many as possible)].

In the proposed scheme, we assign weights to the university according to their value which is known by the position they are placed in QS ranking. We claim, by this way, appropriate recommendation can be made as priority-based weight assignment is used which advocates best-get-best philosophy. But who will decide which recommendation process is performing well? Or which is supposed to be the best? Since, there is no such clear protocol to design a recommender system or to judge a better recommendation process; therefore, we need to evaluate the system, which is, obviously, relative and not absolute. Usually prediction accuracy is considered as the de-facto parameter for the evaluation of recommender systems[38]. It suggests that how accurately the recommendation has been made by the adopted approach? It is obvious that a more accurate system would be desirable by the user. Generally, accuracy is classified as accuracy of ratings predictions, accuracy of usage predictions, and accuracy of rankings of items.

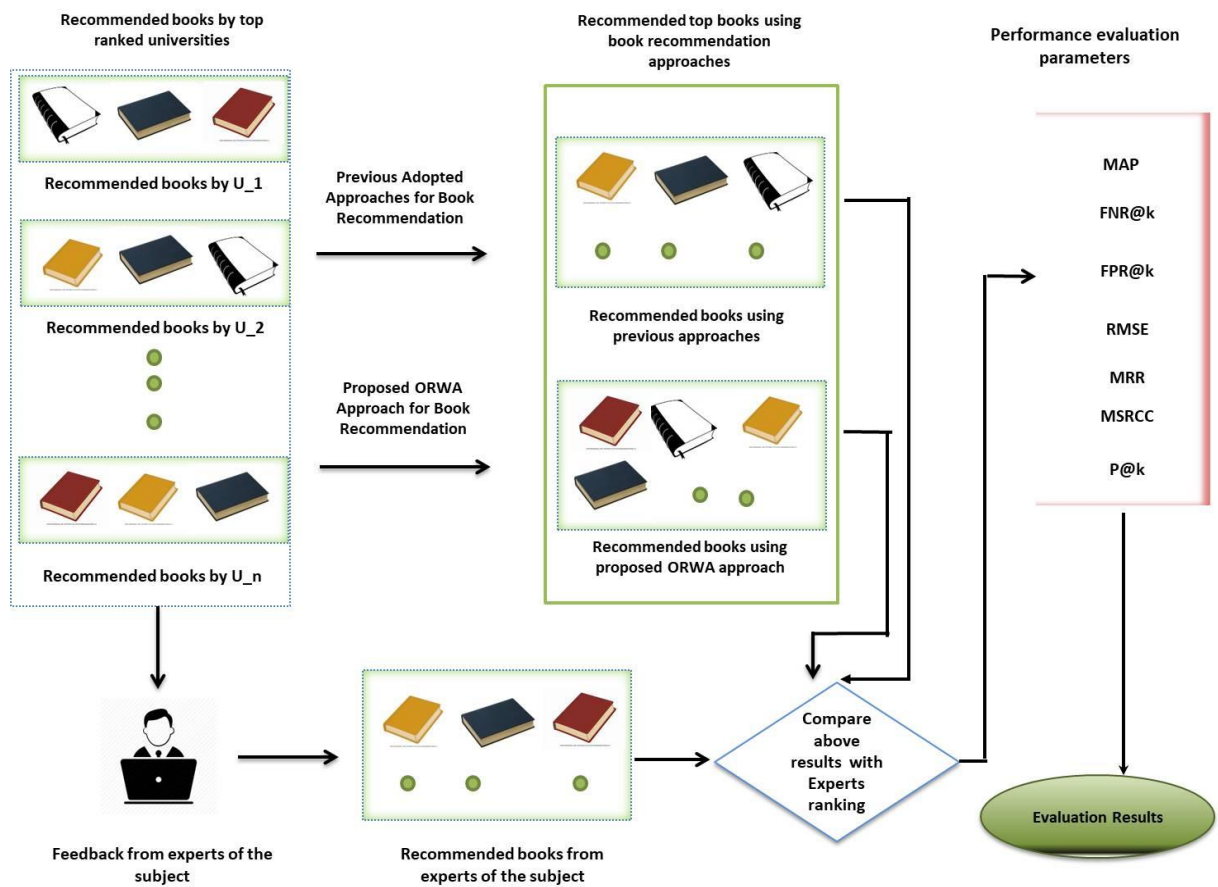

**Figure 4. Performance evaluation scheme for the book recommendation approach**

Since we are advocating human centric aggregation, hence, we have taken experts suggestions for evaluation of the proposed system. These experts are senior academician and Computer Scientists familiar with Indian education systems. We have provided the details of books of respective courses to them. We have adopted evaluation scheme which is based on explicit feedback. The experts' feedbacks for their choice of the books are recorded and these books are ranked, this ranking is compared with the ranking of the books obtained by aforementioned schemes. Thus, the human evaluation from experts would boost the performance evaluation procedure of the adopted recommendation process. Performance evaluation mechanism is diagrammatically presented in Figure 4.

**3.4 Discussion**

The comparison is made on the basis of 8 different parameters. These parameters with their details have been mentioned in section 4.1. Four parameters are used as veracity measures and another four are used as fallacy measures. The value of veracity measures are shown in Figure 5. It is very clear from the graph that OWA (most preferred first) has outperformed the other techniques for all these parameters. The OWA with at least half quantifier, is however, has P@10 and MAP same as that of obtained by OWA (mot preferred first), whereas, the value of modified spearman correlation coefficient and mean reciprocal rank is better for OWA (most preferred first) than OWA (at least half).

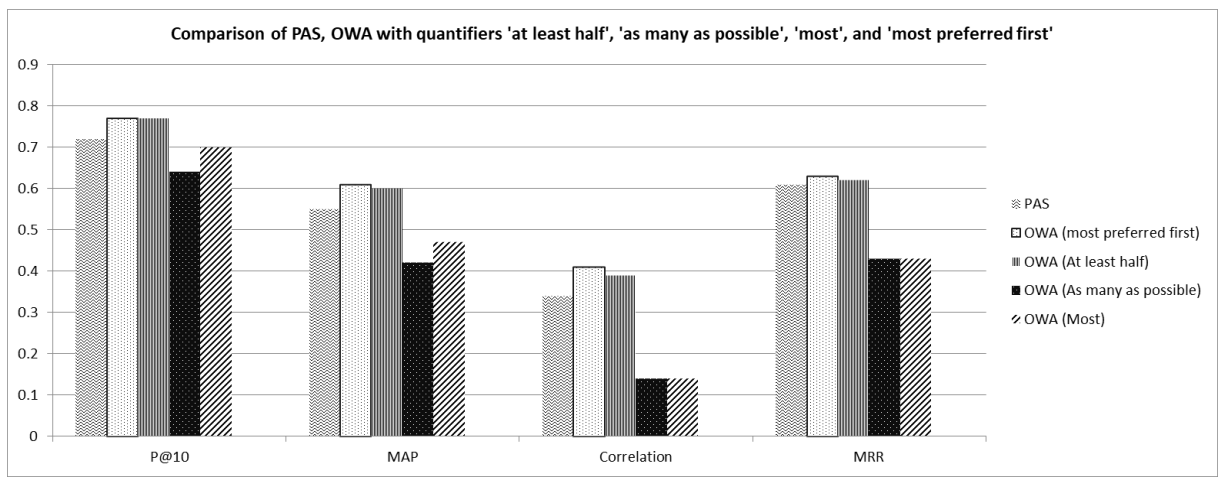

**Figure 5. Comparison of related recommendation approaches for veracity measures including P@10, MAP, MRR and Modified Spearman rank correlation coefficient.**

The improvement in performance of OWA (mot preferred first) for Mean Absolute Error (MAE) with respect to OWA (at least half), (as many as possible) and (most) are 6.66%, 28.20% and 23.28% respectively, whereas with respect to PAS there is an improvement of 8.94%. For Root Mean Square Error (RMSE), 14.64% improved results have been achieved with respect to PAS, whereas, while comparing OWA (mot preferred first) with OWA (at least half), (as many as possible) and (most), it is remarkable that percentages of improvements are 4.24%, 33.46% and 24.55% respectively as shown in

Figure 6. In addition to these results, we can say that with the help of ranked weights, the errors can be reduced which leads to more accurate results.

In figure 7, we can easily notice that the FPR@10 and FNR@10 both have same values for all the approaches. It is because we have full list. For partial list, these two parameters may have different values too. Interestingly the result for OWA (at least half) is same as OWA (mot preferred first), and these results are par from other operators. The false rate is 36% reduced with respect to OWA (as many as possible) and 17.85% improved than PAS. These improvements clearly indicate how powerful OWA can be when treated according to the human-centric situations and how closely they are related to human perception and aggregation.

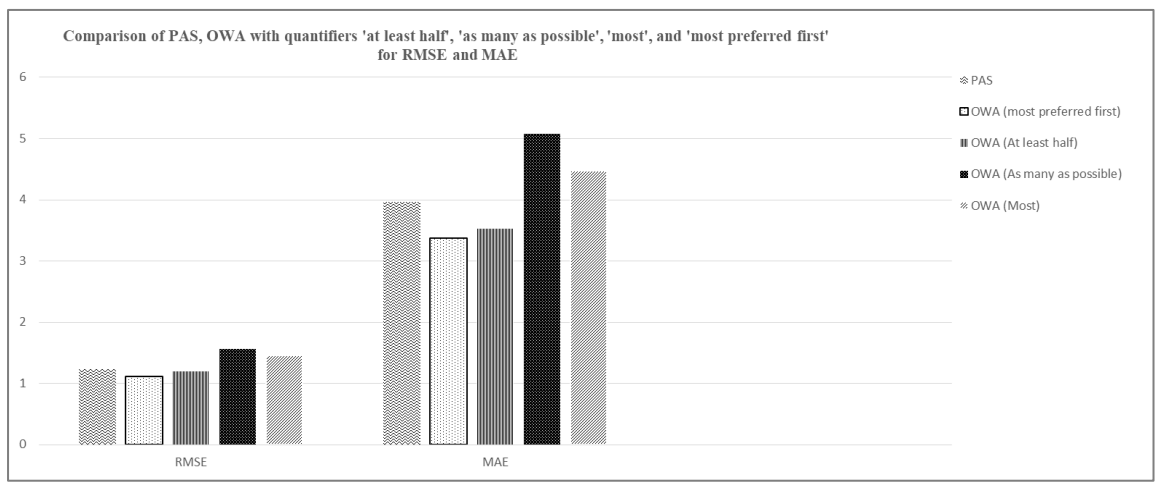

**Figure 6. Comparison of related recommendation approaches for RMSE and MAE**

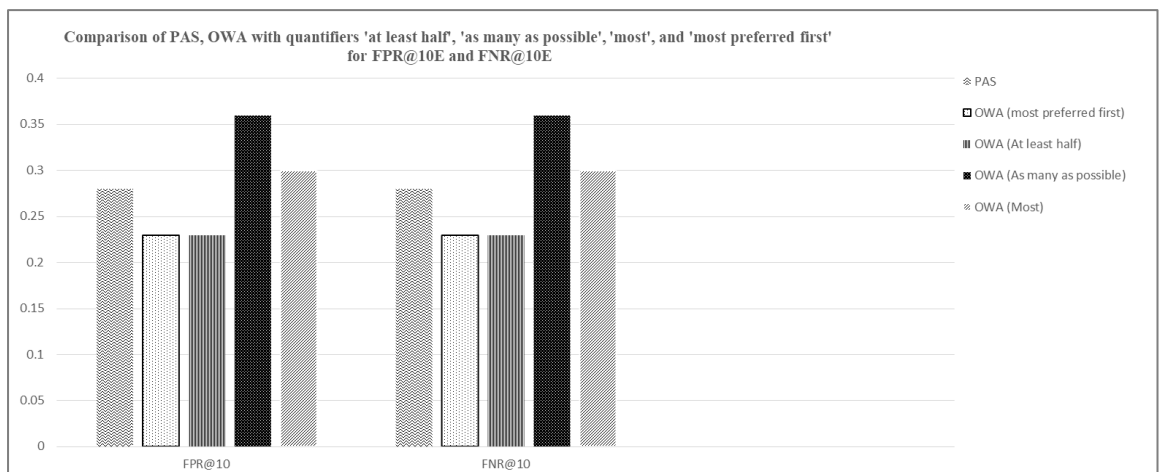

**Figure 7. Comparison of related recommendation approaches for FPR@10 and FNR@10.**

This envisages these quantifiers can be useful for a wide range of applications especially in recommender systems. Also, the human-centric aggregation does not need any prior information of the users' activity, thus it enhance the approach by reducing time and saving storage spaces which are usually required to acquire these knowledge.

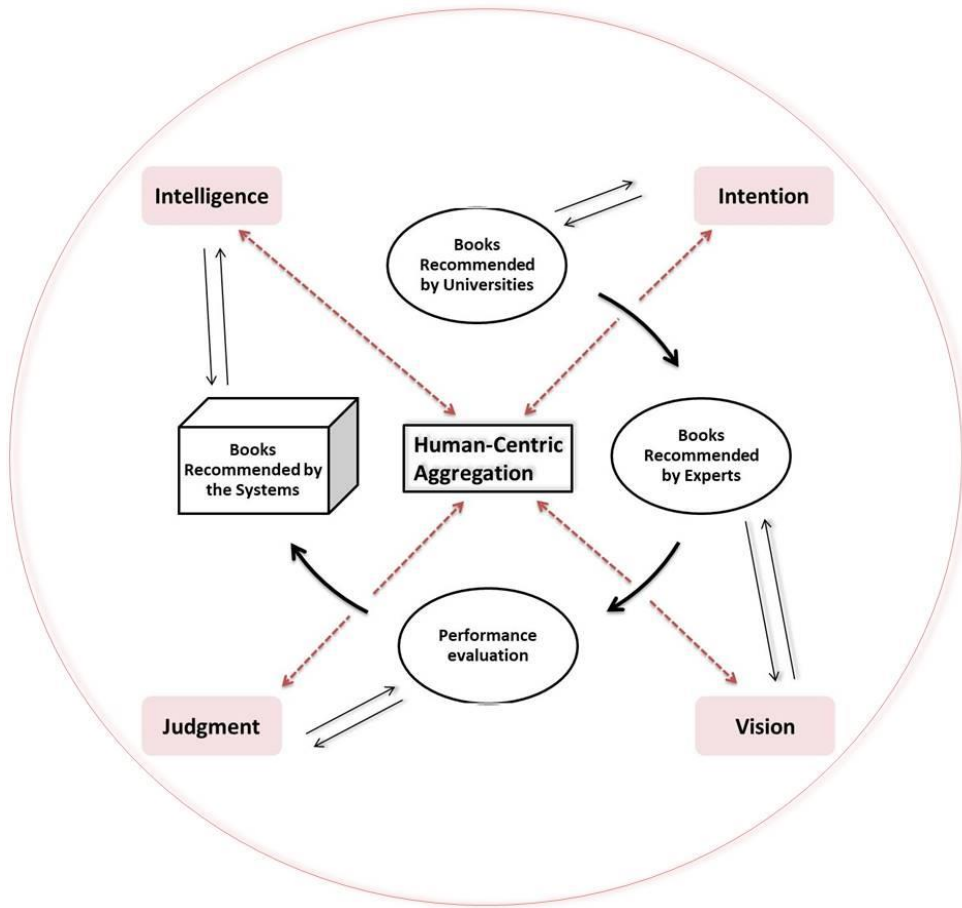

**Figure 8. Book recommendation from the perspective of human-centric aggregation approach**

Further, with the help of OWA and modified weight assignment formula for OWA (mot preferred first), it is observed that the human-centric terms, like intention, intelligence, judgment and vision have been incorporated in this work. The each step in the procedure is based upon human perception, which we say, is human-centric. Since, the ultimate aim of the paper is to apply human-centric aggregation. Initially, intention of ranked universities is taken by involving their recommended books, and then books recommended by experts are considered, which keeps the vision of those experts. Performance evaluation shows the judgement and finally what we recommend is classified as intelligence. Thus different aspects of human perceptions have been assimilated. The concept is diagrammatically demonstrated in Figure 8.

## 4. Conclusion

In this paper, with the proposed methodology, we are intended to incorporate Yager's perspective of linguistic data summaries to explore how human-centric aggregation can be useful in decision making problems, moreover, to know how effective it would be in designing recommender systems. The proposed OWA (mot preferred first), which suggests a modification in weight assignment for OWA operator, has shown improved results with respect to eight parameters over previous OWA with different linguistic quantifiers. The results clearly suggests how powerful human-centric aggregation can be for the above purpose. The presented methodology is envisaged as a platform for future involvement of these aggregation techniques for designing recommender systems. The key in the presented idea is its independence from rating scale, i.e. recommendations have made with more than 78% accuracy, 50% improvement over previous approaches, without exploiting rating scale. One of the considerable contributions of the work is its clarity and also its autonomy from users' prior preferences which provides solution to cold start issues. The proposed approach not only help the university graduates and student community in finding best book for them but also lays a foundation for designing recommender system by using numerical aggregation from human perspectives.

**Conflict of interest:**

On behalf of all authors, the corresponding author states that there is no conflict of interest

**References:**

[1] J. Kacprzyk, R. R. Yager, and J. M. Merigo, "Towards Human-Centric Aggregation via Ordered Weighted Aggregation Operators and Linguistic Data Summaries: A New Perspective on Zadeh's Inspirations," *IEEE Comput. Intell. Mag.*, vol. 14, no. 1, pp. 16–30, Feb. 2019.
[2] G. Ahamad, S. K. Naqvi, and M. M. Beg, "An OWA-Based Model for Talent Enhancement in Cricket," *Int. J. Intell. Syst.*, 2015.
[3] M. Lee, J. Chang, and J. Chen, "Fuzzy Preference Relations in Group Decision Making Problems Based on Ordered Weighted Averaging Operators," vol. 2, no. 1, pp. 11–22, 2014.
[4] R. R. Yager, "On ordered weighted averaging aggregation operators in multicriteria decisionmaking," *IEEE Trans. Syst. Man. Cybern.*, vol. 18, no. 1, pp. 183–190, 1988.
[5] C. Rinner and J. Malczewski, "Web-enabled spatial decision analysis using Ordered Weighted Averaging (OWA)," *J. Geogr. Syst.*, vol. 4, no. 4, pp. 385–403, Dec. 2002.
[6] R. Yager, J. Kacprzyk, and G. Beliakov, *Recent developments in the ordered weighted averaging operators: theory and practice*. 2011.
[7] S. S. Sohail, J. Siddiqui, and R. Ali, "OWA based Book Recommendation Technique," vol. 62, no. Scse, pp. 126–133, 2015.
[8] R. R. Yager, "Using fuzzy measures for modeling human perception of uncertainty in artificial intelligence," *Eng.*

*Appl. Artif. Intell.*, vol. 87, p. 103228, 2020.

[9] R. R. Yager, "OWA aggregation with an uncertainty over the arguments," *Inf. Fusion*, vol. 52, pp. 206–212, 2019.

[10] J. Kacprzyk, S. Zadrożny, "Towards a general and unified characterization of individual and collective choice functions under fuzzy and nonfuzzy preferences and majority via the ordered weighted average operators," International Journal of Intelligent Systems. vol. 4, no. 1, pp. 4-26, Jan 2009.

[11] A. Emrouznejad, M. Marra, "Ordered weighted averaging operators 1988–2014: A citation-based literature survey," International Journal of Intelligent Systems. vol. 29, no. 11, pp. 994-1014, Nov 2014.

[12] C. Rinner, "Exploring Multicriteria Decision Strategies in GIS with Linguistic Quantifiers – A Case Study of Residential Quality Evaluation," vol. 7, pp. 249–268, 2005.

[13] F. Herrera, E. Herrera-Viedma, "A model of consensus in group decision making under linguistic assessments," Fuzzy sets and Systems.vol. 78, no. 1, pp. 73-87, Feb 1996.

[14] R. R. Yager, "Multicriteria decision-making using fuzzy measures," *Cybern. Syst.*, vol. 46, no. 3–4, pp. 150–171, 2015.

[15] J. Kacprzyk, S. Z.-F. S. and Systems, and undefined 2016, "Linguistic summarization of the contents of Web server logs via the Ordered Weighted Averaging (OWA) operators," *Elsevier*.

[16] M. M. S. Beg, "A subjective measure of web search quality," *Inf. Sci. (Ny).*, vol. 169, no. 3, pp. 365–381, 2005.

[17] M. M. S. Beg, "User feedback based enhancement in web search quality," *Inf. Sci. (Ny).*, vol. 170, no. 2, pp. 153–172, 2005.

[18] R. R. Yager, "Fuzzy logic methods in recommender systems," *Fuzzy Sets Syst.*, vol. 136, no. 2, pp. 133–149, 2003.

[19] R. R. Yager, "Intelligent social network analysis using granular computing," *Int. J. Intell. Syst.*, vol. 23, no. 11, pp. 1197–1219, 2008.

[20] B. M. Rasmussen, B. Melgaard, and B. Kristensen, "GIS for decision support-designation of potential wetlands," in *3rd International Conference on Geospatial Information in Agriculture and Forestry*, 2001.

[21] J. Malczewski, "Ordered weighted averaging with fuzzy quantifiers: GIS-based multicriteria evaluation for land-use suitability analysis," International journal of applied earth observation and geoinformation. vol. 8, no. 4, pp. 270-277, Dec 2006.

[22] H. Zabihi *et al.*, "GIS Multi-Criteria Analysis by Ordered Weighted Averaging (OWA): Toward an Integrated Citrus Management Strategy," *Sustainability*, vol. 11, no. 4, p. 1009, Feb. 2019.

[23] S. S. Sohail, J. Siddiqui, and R. Ali, "Book Recommender System using Fuzzy Linguistic Quantifier and Opinion Mining," in *The International Symposium on Intelligent Systems Technologies and Applications*, 2016, pp. 573–583.

[24] R. R. Yager, "Multiple objective decision-making using fuzzy sets," *Int. J. Man. Mach. Stud.*, vol. 9, no. 4, pp. 375–382, 1977.

[25] Z. Xu and R. R. Yager, "Dynamic intuitionistic fuzzy multi-attribute decision making," *Int. J. Approx. Reason.*, vol. 48, no. 1, pp. 246–262, 2008.

[26] R. R. Yager, "OWA aggregation over a continuous interval argument with applications to decision making," *IEEE Trans. Syst. Man, Cybern. Part B*, vol. 34, no. 5, pp. 1952–1963, 2004.

[27] R. R. Yager, "On ordered weighted averaging aggregation operators in multicriteria decisionmaking. IEEE Transactions on systems, Man, and Cybernetics.vol. 18, no. 1, pp. 183-90, Jan 1988.

[28] R. R. Yager, G. Gumrah, and M. Z. Reformat, "Using a web Personal Evaluation Tool–PET for lexicographic multi-criteria service selection," *Knowledge-Based Syst.*, vol. 24, no. 7, pp. 929–942, 2011.

[29] C. Makropoulos, D. B.-E. M. & Software, and undefined 2006, "Spatial ordered weighted averaging: incorporating spatially variable attitude towards risk in spatial multi-criteria decision-making," *Elsevier*.

[30] R. R. Yager, "Multicriteria Decision-Making Using Fuzzy Measures," *Cybern. Syst.*, vol. 46, no. 3–4, pp. 150–171, May 2015.

[31] H. Ishibuchi, K. Nozaki, N. Yamamoto, and H. Tanaka, "Selecting fuzzy if-then rules for classification problems using genetic algorithms," *IEEE Trans. Fuzzy Syst.*, vol. 3, no. 3, pp. 260–270, 1995.

[32] A. Fernandez, F. Herrera, O. Cordon, M. Jose del Jesus, and F. Marcelloni, "Evolutionary Fuzzy Systems for Explainable Artificial Intelligence: Why, When, What for, and Where to?," *IEEE Comput. Intell. Mag.*, vol. 14, no. 1, pp. 69–81, Feb. 2019.

[33] I. Couso, C. Borgelt, E. Hullermeier, and R. Kruse, "Fuzzy Sets in Data Analysis: From Statistical Foundations to Machine Learning," *IEEE Comput. Intell. Mag.*, vol. 14, no. 1, pp. 31–44, Feb. 2019.

[34] S. S. Sohail, J. Siddiqui, and R. Ali, "An OWA-Based Ranking Approach for University Books Recommendation," *Int. J. Intell. Syst.*, vol. 33, no. 2, pp. 396–416, Feb. 2018.

[35] "QS world ranking," *http://www.topuniversities.com/university-rankings/university-subject-rankings/2015/computer-science-information-systems#sorting=rank+region=+country=96+faculty=+stars=false+search=*. [Online]. Available: http://www.topuniversities.com/university-rankings/university-subject-rankings/2015/computer-science-information-systems#sorting=rank+region=+country=96+faculty=+stars=false+search=.

[36] S. S. Sohail, J. Siddiqui, and R. Ali, "A Novel Approach for Book Recommendation using Fuzzy based Aggregation," *Indian J. Sci. Technol.*, vol. 8, no. 1, 2017.

[37] S. S. Sohail, J. Siddiqui, and R. Ali, "A comprehensive approach for the evaluation of recommender systems using implicit feedback," *Int. J. Inf. Technol.*, May 2018.

[38] S. S. Sohail, J. Siddiqui, and R. Ali, "Feature-Based Opinion Mining Approach (FOMA) for Improved Book Recommendation," Arabian Journal for Science and Engineering, vol. 43, no. 12, pp. 8029-8048.

[39] S. S. Sohail, J. Siddiqui, and R. Ali, "Book recommendation system using opinion mining technique," In 2013 international conference on advances in computing, communications and informatics (ICACCI) 2013 Aug 22 (pp.

[39] 1609-1614). IEEE.
[40] S. S. Sohail, J. Siddiqui, and R. Ali, "Recommendation technique using rank based scoring method," In Proceeding of National Conference on Recent Innovations & Advancements in Information Technology (RIAIT-2014) 2014 (pp. 140-146).
[41] Z. Bylinskii, T. Judd, A. Oliva, A. Torralba, and F. Durand, "What Do Different Evaluation Metrics Tell Us About Saliency Models?," IEEE Trans. Pattern Anal. Mach. Intell., vol. 41, no. 3, pp. 740–757, Mar. 2019.
[42] G. Shani and A. Gunawardana, "Evaluating recommendation systems," Recomm. Syst. Handb., pp. 257–298, 2011.